# Models of soft rotators and the theory of a harmonic rotator

*Zahid Zakir* [1]

**Abstract**

The states of a planar oscillator are separated to a vibrational mode, containing a zero-point energy, and a rotational mode without the zero-point energy, but having a conserved angular momentum. On the basis of the analysis of properties of models of rigid and semirigid rotators, the theory of soft rotators is formulated where the harmonic attractive force is balanced only by the centrifugal force. As examples a Coulomb rotator (the Bohr model) and a magneto-harmonic rotator (the Fock-Landau levels) are considered. Disappearance of the radial speed in the model of a magneto-harmonic rotator is taken as a defining property of a pure rotational motion in the harmonic potential. After the exception of energies of the magnetic and spin decompositions, specific to magnetic fields, one turns to a simple and general model of a plane harmonic rotator (circular oscillator without radial speed) where kinetic energy is reduced to the purely rotational energy. Energy levels of the harmonic rotator have the same frequency and are twice degenerate, the energy spectrum is equidistant. In the ground state there is no zero-point energy from rotational modes, and the zero-point energy of vibrational modes can be compensated by spin effects or symmetries of the system. In this case the operators of observables vanish the ground state, i.e. are "strongly" normally ordered. In a chain of harmonic rotators collective rotations around a common axis lead to transverse waves, at quantization of which there appear quasi-particles and holes carrying an angular momentum. In the chain SU (2) appears as a group of symmetry of a rotator.

## Content

[1] *Centre for Theoretical Physics and Astrophyics,* Tashkent, Uzbekistan; zahidzakir@theor-phys.org

# Introduction

At quantization of small *vibrations* of systems they, as a rule, have been reduced to a set of harmonic oscillators having equidistant spectrum and a zero-point energy. By purely formal analogy the oscillatory quantization has been applied to all processes in systems with harmonic potentials. However, a physical sense of frequency is not less often is related by the second kind of periodicity - *rotations*. Formally, the oscillatory quantization has been applied to them also, believing that there something new does not arise in essence.

In fact, in systems with rotational modes there are non-trivial restrictions due to their symmetries. The first restriction is well known and is related by the angular momentum conservation. The second restriction arises from double degeneration of levels and a new discrete symmetry related to it, which do not so widely known.

In the standard courses of quantum mechanics rotators usually associated by *a rigid* rotator of a fixed length corresponding to the constraint $\varphi = r^2 - a^2 \approx 0$, and also with *a semirigid* rotator with a potential $V = \kappa(r-a)^2$ ($\kappa$ - the elasticity coefficient), two ends of which make small vibrations near a fixed length. There is an equilibrium radius $a$ even at lack of rotations, energies grow as $n^2$, frequencies of levels grow also. These models well describe rotational and vibrational levels of molecules and nuclei.

In the present paper another class of rotators, where an attractive force on a plane of rotation is balanced *only* by the centrifugal force, will be considered. They we will call further *as soft rotators*. To them, particularly, concern well-known models of the rotations of a charged particle in the Coulomb field (a *Coulomb* rotator) and in a constant homogeneous magnetic field (a *magneto-harmonic* rotator). In quantum theory these models describe rotational levels of atoms (the Bohr model [1]) and a discrete spectrum of energies of a charge in the magnetic field (the Fock-Landau levels [2,3]).

In the paper a general method of quantization of the soft rotators is proposed and the model *of the harmonic rotator* with a potential of elastic force on the rotation plane is considered. The Lagrangian of the latter model includes the harmonic potential and the purely rotational kinetic term, which corresponds to including into the total Hamiltonian the constraint $\varphi = p_r \approx 0$ excluding a radial momentum $p_r$.

Frequencies of all levels of the harmonic rotator are constant and identical, the energy spectrum is equidistant and in the ground state there is no a zero-point energy of the rotational modes. At compensation of the zero-point energy of the vibrational modes by spin effects or symmetries of the system, the operators of observables vanish at acting to the ground state, i.e. they are normally ordered in a "strong" sense.

In a chain of harmonic rotators collective rotations around a common axis lead to transverse waves which after quantization lead to quasi-particles and holes carrying an angular momentum. As a group of symmetry thus becomes not $O(2)$, as for a set of isolated rotators, but the group $SU(2)$. This follows from the fact that the chain of rotators transfers into itself not at rotation to the angle $2\pi$, but to the angle $4\pi$ when it is possible to untangle a chain without changing of edges (the Dirac method). As the result, in the chain the quanta in a spinor representation of the rotation group can appear.

At inserting of interactions of quanta of the harmonic rotator by perturbation theory, the operators in a Hamiltonian and currents are "strongly" normally ordered. Such ordering can be viewed as a part of the recipe of transition from the oscillatory to rotatory quantization.

In Section 1 of the paper a short review of using in practice models of rotators is presented, in Section 2 the theory of the harmonic rotator and the chains of harmonic rotators are constructed.

## 1. Practically using models of rotators

### 1.1. Rigid rotator

Two particles anchored on ends of a rigid rod of length $r = a$, rotating around the center of mass with frequency $\omega$, form *a rigid rotator*. The standard physical example of the rigid rotator is a diatomic molecule, where the difference of energies of vibrational levels sufficiently exceeds the difference of rotational energy levels. Rotational levels, therefore, are placed between the vibrational ones and, as a first approximation, the energy of vibrations can be considered without the account of rotations. Then to each such level the energy of rotations of two centers of oscillation can be added by supposing the molecule as a rigid one.

The wave equation of the system of two particles with a reduced mass $m$

$$-\frac{\hbar^2}{2m}\nabla_r^2\psi - \frac{\hbar^2}{2mr^2}\frac{\partial^2\psi}{\partial\theta^2} + E\psi = 0 \tag{1}$$

at the approximation $r \approx a$, when energy associated by the rotation of two centers, is reduced to

$$E\psi = -\frac{\hbar^2}{2ma^2}\frac{\partial^2\psi}{\partial\theta^2}. \tag{2}$$

The angular momentum and the Hamiltonian of the rigid rotator are:

$$M = -i\hbar\frac{\partial}{\partial\theta}, \quad H = \frac{M^2}{2ma^2}, \tag{3}$$

the angular variable $\theta$ is a cyclical variable, and thus corresponding generalized momentum, the angular momentum, is conserved:

$$M = mr^2\dot\theta = mr^2\omega = const. \tag{4}$$

Solutions $\psi_n$ of the wave equation on a rotation plane contain frequencies of $\omega_n$ both rotation directions:

$$\psi_n(\theta,t) = \frac{1}{\sqrt{2\pi}}\exp\left[(iE_n t + iM_n\theta)/\hbar\right], \tag{5}$$

$$E_n = \frac{M_n^2}{2ma^2} = M_n\omega_n = n\hbar\cdot\omega_n,$$

$$M_n = n\hbar, \quad n = 0, \pm 1,... \tag{6}$$

$$\omega_n = n\omega', \quad \omega' = \hbar/2ma^2.$$

Thus, both the angular momentum and the frequency of the rigid rotator grow proportionally $n$. As the result, the energy of levels grow as $n^2$ and thus the energy spectrum is similar to the spectrum of a particle in a potential well.

However, in the potential well there is a lower level $n = 1$ with nonzero energy that is required by the uncertainty relations, whereas the spectrum of the rigid rotator (6) begins from $n = 0$ which leads to the energy $E_n = 0$. This paradox in the form of the explicit contradiction with the uncertainty relations can be solved if one turns to the initial wave equation (1) for the whole system and takes into account that the *rigid rotator approximation* $r \approx a$ *inapplicable to the state* $n = 0$. In this state the rotational contribution with $1/r^2$ disappears because of vanishing of the angular momentum and the neglecting of the radial kinetic term is impossible. Therefore the equation (1) turns not to (2), but to:

$$-\frac{\hbar^2}{2m}\nabla_r^2 \psi_0 + E_0 \psi_0 = 0 \qquad (7)$$

and solutions of this equation contain the contribution of radial (*vibrational*) fluctuations with $\delta E_0 > 0$ in accordance with the uncertainty relations. This feature of the spectrum of rotators at lack of rotations one must take into account for all types of rotators in which radial modes do not forbidden by the symmetries of a system.

The double degeneration of the rotational levels generates a discrete symmetry with a new conserved quantity – a chiral charge $Q$ equal to the projection of the angular momentum on the normal to the rotation plane $\mathbf{n}_z$:

$$Q_n = \mathbf{n}_z \cdot \mathbf{M}_n. \qquad (8)$$

A sign on the angular momentum, according to Eq. (4), is defined by a sign on frequency of rotation, whereas the sign on $Q$ depends on a sign on the normal also.

The physical sense of $Q$ becomes clear if we consider the *rotators evolving backward in time* which is formally admissible because of the time reversal symmetry. Thus, rotation in the positive direction changes on reverse, while a direction of the angular momentum and a sign on the chiral charge remain unchanged and positive. But, since the frequency changes a sign, according to Eq. (6) the sign on energy changes also. These *states of the negative energy going backward in time* then we interpret as the *charge-conjugate states of the positive energy and frequency, going forward in time* (*the crossing symmetry*).

Thus, the properties of the energy spectrum of the rigid rotator, common for other types of rotators also, are that:

1) the states are *doubly degenerate* for two directions of rotation;

2) the energies are proportional to the angular momentum and frequency: $E_n = M_n \omega_n$;

3) in the ground state $n = 0$ there is no *zero-point energy* and the *zero-point angular momentum*;

4) in the first nontrivial state $n = \pm 1$ momentum is of order $p \sim \hbar/a$;

5) there is the chiral symmetry of states going forward or backward in time, the corresponding chiral charge is conserved $Q = const.$;

6) the rotational energy levels are symmetrical under zero $E_0 = 0$.

### 1.2. Semirigid vibrating rotator

In the case of a diatomic molecule, as an example of a real physical system, the deviations from the idealized concept of a rigid rotator appear as that when there is no a rotational mode, nevertheless there are zero-point vibrations of atoms about equilibrium position at a lowest vibrational level. Therefore, for completeness of the description it is necessary to consider the vibrational degrees of freedom of the system also.

Let ends of a rod are not fixed rigidly and two particles of the rotator make small vibrations about equilibrium positions $r = a \pm \delta r$. The small vibrations can be described by means the harmonic oscillator potential $\kappa(r-a)^2/2$, where $\kappa$ is the elasticity coefficient and we come to the model *of a semirigid vibrating rotator*. In the Hamiltonian the rotational kinetic term behaves as a potential and it can be considered as a part of an effective potential energy $U_{\mathit{eff}}$:

$$H = \frac{p_r^2}{2m} + U_{\mathit{eff}}(r)$$

$$U_{\mathit{eff}}(r) = \frac{M^2}{2mr^2} + \frac{\kappa}{2}(r-a)^2. \qquad (9)$$

The equilibrium position corresponds to a minimum of total energy that is defined now not by a former point $r \approx a$, but by the minimum of the effective potential $U_{\mathit{eff}}(r_0)$ placed at a balance point of the elastic and centrifugal forces:

$$\frac{M^2}{mr_0^3} = \kappa|r_0 - a|. \qquad (10)$$

At the first approximation it is possible to neglect interaction of the vibrations and rotations and then the energy levels are reduced to the sum of rotational and vibrational energies with two types of frequencies:

$$E_n = n_\theta \hbar \cdot \omega_{n_\theta} + \left(n_r + \frac{1}{2}\right)\hbar\omega_0$$

$$\omega_{n_\theta} = \frac{n_\theta \hbar}{2ma^2}, \quad \omega_0 = \sqrt{\frac{\kappa}{m}}. \qquad (11)$$

Corrections to Eq. (11) we can find by expansion of the rotational term in Eq. (9) under small deviations from $a$.

The model of a semirigid vibrating rotator at the first approximation well describes vibrational and rotational modes of diatomic molecules and some deformed nuclei.

In the present system most important and new with respect to a rigid rotator property of the present system is the equilibrium condition of the centrifugal and elastic forces in Eq. (10). In case of soft rotators this condition takes on essential significance, which is well-known already for the Coulomb rotator which will be considered below.

### 1.3. The Coulomb rotator (Bohr model)

The first, most known and successful application of a soft rotator in quantum theory was *the Coulomb rotator*, where the centrifugal force is balanced by the Coulomb attraction. This type of rotator has been used in Bohr's model for the hydrogen-like atoms [1]. In this model we will be interested by those its properties which can be useful at further consideration of a harmonic rotator.

The Hamiltonian of the model generally looks like

$$H = \frac{p_r^2}{2m} + \frac{M^2}{2mr^2} - \frac{Ze^2}{r}, \qquad (12)$$

and the equations of motion for a radial variable are:

$$m\dot{r} = p_r, \quad \dot{p}_r = \frac{\partial H}{\partial r} = -\frac{M^2}{mr^3} + \frac{Ze^2}{r^2}. \qquad (13)$$

The effective potential energy

$$U_{eff}(r) = \frac{M^2}{2mr^2} - \frac{Ze^2}{r} \quad (14)$$

and also the total energy has minima at a balance point of the centrifugal and Coulomb forces:

$$\frac{M_n^2}{mr_n^3} - \frac{Ze^2}{r_n^2} = 0. \quad (15)$$

This condition gives equilibrium radius and expression for energy of orbits with purely rotational kinetic term:

$$r_n = \frac{M_n^2}{mZe^2}, \quad E_n = -\frac{mZ^2e^4}{2M_n^2}, \quad (16)$$

which correspond to the levels of the Coulomb rotator. Then quantization of the angular momentum for these levels on the rotation plane gives levels of energy with a "magnetic" quantum number $n_\varphi$:

$$M_n = n\hbar = (n_\varphi + 1)\hbar, \quad n_\varphi = 0, \pm 1, \ldots \quad (17)$$

$$E_n = -M_n \omega_n = -n\hbar \omega_n, \quad \omega_n = \frac{mZ^2e^4}{2\hbar^3 n^3}. \quad (18)$$

The shift on unit in Eq. (17) is related by the radial fluctuations in the absence of rotation, as it was pointed out in the case of a rigid rotator. The levels in Eq. (18) coincide with the results of the solution of the Schrödinger equation when $n_r = 0$, the angular quantum number is equal to the "magnetic" quantum number $l = n_\varphi$, and the principal quantum number is $n = l + 1 = n_\varphi + 1$. As we see, at growing $n$ the rotation frequencies of levels of the Coulomb rotator decrease as $n^{-3}$ and energies as $n^{-2}$.

At transitions between the purely rotational levels the frequency of emitted photons $\omega_{nn'}$ and their energies, thus, are equal to:

$$\Delta E_{nn'} = \hbar \omega_{nn'} = \hbar(n'\omega_{n'} - n\omega_n) = \frac{mZ^2e^4}{2\hbar^3}\left(\frac{1}{n'^2} - \frac{1}{n^2}\right). \quad (19)$$

As we see, the frequencies of photons depend on not only a difference of rotation frequencies of electrons at two levels, but a difference of angular momenta at these levels also. This feature explains the known paradox of the Bohr model when the frequencies of emitted photons differ from the rotation frequencies of electrons.

### 1.4. Magneto-harmonic rotator (Fock-Landau levels)

Solutions of the Schrödinger equation for a charged particle in a constant homogeneous magnetic field $\mathbf{H} = (0, 0, H_z)$ and corresponding levels of energy have been found by Fock in 1928 [2] which showed that formally they are reduced to well-known formulas for harmonic oscillator. Landau in 1930 obtained them in another calibration and applied to the theory of diamagnetism [3]. Then the discrete levels of energy of a charge in the constant magnetic field have been widely applied (it is accepted to name them as Landau levels, but since Fock published the solution two years earlier, we will name further they as the Fock-Landau levels).

The physically most natural choice of gauge for the vector-potential for the constant and homogeneous magnetic field reads:

$$\mathbf{A} = \frac{1}{2}[\mathbf{H} \times \mathbf{r}]. \quad (20)$$

The Schrödinger equation with this potential:

$$-\frac{\hbar^2}{2m}\nabla_r^2 \psi + \left(E + \frac{1}{2}M\omega_H - \frac{M^2}{2mr^2} - \frac{m\omega_H^2}{8}r^2\right)\psi = 0, \quad (21)$$

gives the energy levels:

$$E_{n_r,n_\theta} = \frac{1}{2}\hbar|\omega_H|(2n_r + |n_\theta| + 1) + \frac{1}{2}\hbar\omega_H n_\theta + \frac{p_z^2}{2m}. \quad (22)$$

Here $p_z$ is momentum along the magnetic field, $r^2 = x^2 + y^2$ and $\omega_H$ is the cyclotron frequency:

$$\omega_H = \frac{eH_z}{mc}. \quad (23)$$

Since in the classical case motion in a magnetic field on a rotation plane, symmetrical under both axes, also happens *only* on circular orbits, in the quantum case there are no radial excitations ($n_r = 0$). The energy of zero-point radial vibrations of the charged particle is transformed by the magnetic field to the energy of "zero-point rotations" without changing a value of this energy and consequently the lowest level on a rotation plane (without energy of motion along the field) is reduced to $E_0 = \hbar|\omega_H|/2$. As the result the levels of Eq. (22) take a form:

$$E_{n_\theta \pm} = \frac{1}{2}\hbar|\omega_H|(|n_\theta| + n_\theta + 1) + \frac{p_z^2}{2m}. \quad (24)$$

Here the sign $n_\theta$ corresponds to a sign on angular frequency $\omega_H$ (23), depending (at given direction of magnetic field) on a sign on charge of the particle. Thus, in Eq. (24) for each sign $\omega_H$ (or the charge sign) there contribute to the energy only levels with $\omega_H n_\theta > 0$ and contributions of levels with $\omega_H n_\theta < 0$ disappear:

$$E_{n_\theta} = \begin{cases} \hbar|\omega_H|(|n_\theta| + 1/2), & \omega_H n_\theta > 0, \\ \hbar|\omega_H|/2 & \omega_H n_\theta < 0. \end{cases} \quad (25)$$

At taking into account the spin of electron $s$ the ground state split on spin levels:

$$E_{0,s} = \hbar|\omega_H|\left(s + \frac{1}{2}\right). \quad (26)$$

As the result, for electron with $s = \pm 1/2$ the lowest energy shifts to zero $E_{0,-1/2} = 0$, the next level $E_{0,+1/2} = \hbar|\omega_H|$ and all other levels are double degenerate.

Thus, the problem of charged particle's motion in the constant homogeneous magnetic field, in fact, is the problem of a *magneto-harmonic rotator* with the harmonic potential balancing centrifugal force, and also with energies of magnetic and spin splitting. This model is widely applied practically in all systems of condensed media physics and astrophysics where there are quantum effects in magnetic fields.

Thus, the energy spectrum of the magneto-harmonic rotator, on the one hand, has set of properties common with the harmonic oscillator spectrum:
1) frequencies of all levels are *identical*;
2) the dependence of energy on a stretching is a *parabola*;

   3) energy levels are *equidistant*;
   4) there is the *zero-point energy* in the ground state due to radial modes.

On the other hand, there is also a set of differences from harmonic oscillator:

   1) states *are double degenerate* for two directions of rotation;
   2) energy is proportional to an angular momentum and frequency: $E_n = M_n \omega_n$;
   3) rotational modes at $n_\theta = 0$ *do not contribute* to the zero-point energy, and at the spin splitting the zero-point energy from vibrating modes may be compensated;
   4) energy levels are symmetrical under zero $E_0 = 0$;
   5) there is a chiral symmetry between states going forward or backward in time;
   6) corresponding chiral charge $Q = \mathbf{n}_z \mathbf{M}$ is conserved.

Main property of motion in the homogeneous magnetic field is that at switching-on the field a value of speed of a charge after that moment *becomes entirely tangential* component of speed $\mathrm{v} = (0, r\dot\theta)$, *the radial component of speed disappears* $\dot r = 0$ and further latter does not appear. This means that in the classical case in the magnetic field there are no the radial vibrations and there are pure rotations only. Therefore, though one can use a formal analogy to harmonic oscillator, but only by taking into account the constraint $\dot r = 0$.

## 2. Model of a harmonic rotator

### 2.1. Classical harmonic rotator

The vanishing of a radial component of speed, appeared in the model of a magneto-harmonic rotator, can be taken further as a defining property of a new class of motions in the systems with harmonic potentials. If we eliminate from the model of a magneto-harmonic rotator the specific for magnetic field contributions to energy, such as magnetic and spin splitting, we come to simpler and more general model - *a harmonic rotator* model.

This model, in other hand, is only a special case of the models of soft rotators, the Lagrangian of which includes purely rotational kinetic term and an attraction potential $V(r)$ on the rotation plane:

$$L = \frac{1}{2} m r^2 \dot\theta^2 - V(r). \tag{27}$$

In the first part of the paper practically using models of soft rotators have been considered and thus in the present second part we will consider mainly a new model of the harmonic rotator in case of which the specific properties and perspectives of applications of soft rotators will be studied.

In the model of a harmonic rotator, hence, it is enough to leave the elastic force potential $V(r) = \kappa r^2 / 2$ on the rotation plane and to restrict by a case when kinetic term is reduced to the rotational energy only. The harmonic rotator has common features and differences both with respect to the linear harmonic oscillator, and with respect to usually applied models of a rigid rotator and a semirigid rotator with two centers of vibration.

Common features and differences of the harmonic rotator's energy from the linear oscillator's energy are clearly evident at considering a classical particle of mass $m$, at the initial moment placed at the origin of coordinates and bounded to this point by an elastic force. Let then other free particle of the same mass scatters with this particle and transfers to it kinetic energy $mv_r^2 / 2 = \kappa a^2 / 2$. The bounded particle then deviates from the center up to the point $r = a$, where it stops one instant, and its kinetic energy fully transforms to potential energy of the elastic force.

Then, if there is no exterior influence, the particle returns to center with former kinetic energy and will make further radial vibrations as a linear harmonic oscillator with total energy $E = ma^2\omega^2/2 = \kappa a^2/2$, which reduces to kinetic energy at $r = 0$ and to potential energy at $r = a$.

If at stopping at $r = a$ once more free particle of the same kinetic energy $mv_\perp^2/2 = \kappa a^2/2$ scatters with the bounded particle, but with a momentum directed perpendicularly to radius, the bounded particle turns to rotate around center on a circle of radius $r = a$ and will represent a harmonic rotator. The total energy of such harmonic rotator $E = \kappa a^2 = ma^2\omega^2$ is the sum of kinetic and potential energies (which are constant and equal each other) and twice more total energy of the linear harmonic vibrations. A plot of dependence of the total energy of the harmonic rotator from the deviation $a$ is a parabola with center at $r = 0$.

Thus, the Lagrangian of the harmonic rotator contains a pure rotational kinetic term and the harmonic potential:

$$L = \frac{1}{2}mr^2\dot{\theta}^2 - \frac{1}{2}\kappa r^2. \tag{28}$$

There no a radial speed and $r$ ceases to be a dynamic variable. If to rewrite a Lagrangian (28) in the form

$$L = r^2 \frac{m}{2}\left(\dot{\theta}^2 - \omega_\pm^2\right), \quad \omega_\pm = \pm\omega = \pm\sqrt{\kappa/m}, \tag{29}$$

we see that $r^2$ turns into the Lagrange multiplier, and a constraint $\dot{\theta} = \pm\omega$ corresponding to this multiplier leads to independence of angular speed (frequency of rotation) on radius. The angular momentum conservation fixes the radius:

$$r^2 = \frac{M}{m\omega}, \tag{30}$$

and energy:

$$E_n = m\omega r^2 = M\omega. \tag{31}$$

Thus, energy of the harmonic rotator, though is equal to energy of a circular harmonic oscillator, but is proportional to the angular momentum.

Since the Lagrangian (29) was reduced to the pure constraint, the Hamiltonian is reduced to a product of momentum to speed of the dynamical variable $\theta(t)$ plus the same constraint, which is conserved in time and can be trivially solved:

$$H = p_\theta\dot{\theta} - L = M\dot{\theta} - r^2\frac{m}{2}\left(\dot{\theta}^2 - \omega_\pm^2\right) = M\dot{\theta}\Big|_{\dot{\theta}=\omega_\pm} = M_\pm\omega_\pm. \tag{32}$$

Thus, the Hamiltonian of the harmonic rotator appears as extremely simple and linearly depending both on the angular momentum and frequency.

In this case the constraints are easily solved and there is no necessity for application of full formalism of the canonical quantization of systems with constraints. Nevertheless, the consideration of the canonical quantization of the harmonic rotator is useful as for to complete the picture, and for the further applications in the presence of other potentials.

Let a nonrelativistic particle of the reduced mass $m$ moves on a plane in the harmonic potential, but with a constraint $\varphi(r, p_r)$ to the radial variables. In case of harmonic rotator frequency of rotation, as well as frequency of vibration for harmonic oscillator, is constant, and the angular momentum and energy levels are defined by a value of deviation. Therefore unlike a rigid rotator, $r$ is not constant, but is a dynamically fixed parameter. Common for all states of the harmonic rotator have a purely rotational character of motion around center that means absence of radial speed or radial momentum. Therefore, initially there is one a primary constraint providing the vanishing of a radial momentum:

$$\varphi(r, p_r) = p_r \approx 0. \tag{33}$$

At the standard canonical quantization starting with the Hamiltonian of the particle on the plane with harmonic potential

$$H = \frac{p_r^2}{2m} + \frac{M^2}{2mr^2} + \frac{1}{2}kr^2. \tag{34}$$

a generalized Hamiltonian with constraint (33) should be constructed:

$$\tilde{H} = H + \lambda p_r, \tag{35}$$

where $\lambda(t)$ is the Lagrange multiplier and the Lagrangian is defined as $\tilde{L} = p_r \dot{r} + M\dot{\theta} - \tilde{H}$.

The equation of motion on the radial coordinate on the surface of constraint $p_r \approx 0$ then is the equation of balance between the elastic and centrifugal forces:

$$-\frac{M^2}{mr^3} + kr = 0. \tag{36}$$

Then the using of the expression for $M$ gives to constancy of the angular frequency $\dot{\theta} = \pm\omega = \pm\sqrt{k/m}$.

The Poisson bracket of the constraint (33) with the Hamiltonian also is reduced to the equilibrium condition (36) and also disappears on the surface of constraints:

$$\{\tilde{H}, \varphi\} = \{H, p_r\} = \frac{1}{2m}\left\{\frac{M^2}{r^2} + kr^2, p_r\right\} = \frac{1}{m}\left(-\frac{M^2}{r^3} + kr\right) \approx 0. \tag{37}$$

As the result, the secondary constraints do not arise and further, eliminating $r$ from the expression for energy with the help (36), we come to (32).

As we see, the result of standard canonical quantization from the physical point of view completely coincides with the above presented simple method when $r^2$ enters into the Lagrangian as a Lagrange multiplier and there is no momentum for $r$.

The case of *a rigid* rotator (see section 1.1) corresponds to the primary constraint $\varphi = r^2 - a^2$. Here after the resolution of the constraint $r$ becomes equal to the constant $a$ fixed externally, its momentum disappears, and the potential is degenerate in an additive constant which can be dropped. As the result, the Lagrangian on a surface of constraint reduces to the standard form where there is only rotational kinetic term for the angular variable $M^2/2ma^2$, and, due to a fixed length, transitions between energy levels happen through angular frequency changings only.

## 2.2. Quantization of a harmonic rotator

The proper values $M_n$ and the proper functions $\psi_n(\theta)$ of the angular momentum on the rotation plane are:

$$-i\hbar \frac{\partial \psi_n}{\partial \theta} = M_n \psi_n = n\hbar \psi_n, \tag{38}$$

$$\psi_n(\theta) = \frac{1}{\sqrt{2\pi}} e^{in\theta}. \tag{39}$$

Therefore the appropriate Schrödinger equation for the harmonic rotator:

$$E\psi = -\frac{\hbar^2}{2mr^2} \frac{\partial^2 \psi}{\partial \theta^2} + \frac{1}{2}\kappa r^2 \psi \tag{40}$$

can be simply solved and the results coincide with the results of above presented simple derivation. Really, a total energy, according to (40) and with adding of energy of zero-point radial fluctuations $E_0 = \hbar\omega/2$, is:

$$E_n = \frac{n^2\hbar^2}{2mr_n^2} + \frac{1}{2}\kappa r_n^2 + E_0 \tag{41}$$

and, in a combination with proper values of the angular momentum (38), allows us to express $r_n^2$ and obtain the expression for the energy levels:

$$r_n^2 = \frac{M_n}{m\omega} = \frac{n\hbar}{m\omega}, \tag{42}$$

$$E_n = |M_n|\omega + E_0 = \hbar\omega(|n|+1) = |Q_n|\omega, \tag{43}$$

where $Q_n = -(M_n + 1)$ is a "charge" operator. As we see, the *energy spectrum of the harmonic rotator practically coincides with the spectrum of the linear harmonic oscillator*, but with that difference that the zero-point energy of the ground state does not concern to rotational degrees of freedom. Quantization of energy of the harmonic rotator thereby is reduced to quantization of the angular momentum and practically is reduced to the specrum of the "charge" operator.

At comparing of harmonic rotator's energy spectrum with harmonic oscillator's energy spectrum there are three common properties and three distinctions. The common ones consist that:

1) frequencies of all energy levels are *identical*;
2) a plot of dependence of energy from radius is a *parabola*;
3) energy levels are *equidistant*.

Differences consist that in the harmonic rotator spectrum:

1) energy levels are *doubly degenerate*;
2) in the ground state *there is no zero-point energy* from purely rotational modes;
3) energy levels *are proportional* to the angular momentum and frequency.

The properties of the harmonic rotator are characteristic for many systems with complex variables, particularly, relativistic fields, and consequently it is natural their quantization perform on the basis of the harmonic rotator model. In the systems with charge-conjugation symmetry instead of the angular momentum figures a charge operator. A key symmetry of such systems is a symmetry between the charge-conjugate states of the same energy.

### 2.3. Quantization of rotating chain of harmonic rotators

The spectrum of a harmonic rotator is equidistant, can start from zero and is double degenerate that is of interest at quantization of relativistic fields with gauge and charge-conjugation symmetries. But the fields should be similar to chains of rotators of very small step and thus we turn to consider finite step chains.

Let there is a set of $N$ identical harmonic rotators, each formed by a pair of elastically connected and rotating on parallel planes $(x, y, z_i)$, $(i = 1,...,N)$ pairs of particles $A_i$ and $B_i$, and let the centers of inertia of these rotators allocated with step $d$ along the common axis of rotation $z$. At identical angular speed $\omega_0$ and identical initial conditions all rotators of the set rotate coherently and, because of angular momentum conservation, each of them remains on own rotation plane. Thus, energy of the system is equal to

$$E_N = N_+ M_+ \omega_{0+} + N_- M_- \omega_{0-} + E_0, \tag{44}$$

where $M_\pm$ is the angular momentum of one a rotator.

Now let's connect by means of elastic force $\kappa(r_{i+1} - r_i)$ each of two particles of any rotator to similar particles of two adjacent rotators ($A_i$ with $A_{i\pm1}$ and $B_i$ with $B_{i\pm1}$), by forming a double one-dimensional chain of $2N$ particles. In the continuous limit this double chain forms two massive strings connected by an elastic film of small width.

Frequency of rotation of each harmonic rotator $\omega_0 = \sqrt{\kappa/m}$ is constant and the connections with neighbors do not change it. They only lead to change of the angular momentum at an additional stretching on the rotation plane, thus because of the angular momentum conservation and weakness of connections with neighbors each rotator remains on its rotation plane.

Let now the enclosed exterior force transfers to one of rotators in the chain an additional angular momentum and appropriate energy. Then that rotator act on two nearest neighbors, transferring them a part of the angular momentum and energy, and those then act on subsequent and the angular momentum will be propagated on the chain. All chain thus continues to rotate synchronously, only a small excess stretching starts to be moved on the chain and in the system *a transverse wave of stretching* of rotators with changing and reversing of their angular momenta will be propagated.

The wave length $\lambda = l'd$ is defined by a number of rotators $l'$ covered by one period of this wave. The appropriate wave vector $\mathbf{k}_z = 2\pi \mathbf{n}_z / \lambda$ is directed along $z$ axis and will set further momentum of the wave quantum. "Longitudinal" frequency $\omega_z = v_z k_z$ of propagating wave is product for $k_z$ and speed of wave $v_z$. Thus, two crests of waves on opposite ends of rotators rotate on the plane $(x, y)$ with the common angular frequency $\omega_0$, being propagated along $z$ with speed $v_z$, so that their way in space are two helicoid curves round $z$ axis.

In the quantum theory where each of rotators has a quantized angular momentum and appropriate level of energy, the passage of the wave consists in passage on of a rotator to upper or lower level and reverse. If a unit angular momentum has been transferred to one of rotators by exterior force, at propagation of this perturbation on the chain each of chain's rotators in turn passes to *higher level* and reverse, and this leads to a *quasi-particle* propagation. At perturbation of the chain by take away a unit angular momentum from one of rotators, other chain's rotators in turn passes to *lower level* and reverse, and this leads to a *hole* propagation. If to the same rotator of the chain from one hand the quasi-particle arrives, and on the other hand simultaneously arrives a hole, the angular momentum will not changes and thus there occurs a *recombination* of these perturbations.

In a system of unconnected with each other, allocated along the common axis of rotation and coherently rotating rotators there is *a rotational symmetry* $O(2)$ in space $(x_i, y_i)$ of two degrees of freedom. However, when rotators form an elastically bounded chain, the connections of the rotators with each other leads to a new effect of topological nature. The system passes in itself not at turnaround on $2\pi$ separately of any rotator in a chain, but at its turnaround on $4\pi$, and thus a symmetry group at rotations becomes $SU(2)$. Really, turn on $2\pi$ one of rotators without turn of boundaries of the chain changes connections with neighbors and, obviously, there appears another state. At turn of this rotator on $4\pi$ the chain can be untangled (the Dirac method) and it is returned in an initial state without change of boundaries of the chain and without change of orientation of the turned rotator. This means that in the chain of the bounded rotators there appear naturally quanta in spinor representations of the rotation group.

More detailed consideration of the quantum theory of such chains and consequences of their symmetry properties will be presented in subsequent papers.

## Conclusion

Thus, the absence of a radial component of the speed, initially appeared in the pure form in the magneto-harmonic rotator, is a defining property of the soft rotators, in particular, the harmonic rotator. On the experience of construction of using in practice models of soft rotators, this property allowed us to formulate their general theory and, in particular, the quantum theory of the harmonic rotator.

Quantization of waves at collective rotations of a one-dimensional chain of harmonic rotators allows us to model of fields with rotational symmetry and, probably, with spinor degrees of freedom.

In the following paper the rotational quantization of systems, where frequencies of quanta concern to angular frequencies of harmonic rotators or field vectors, will be developed. There a new one will be not only naturalness of inserting into the theory of angular and spin observables, but also possibility to eliminate a zero-point energy and a zero-point charge of the ground state as it happens at spin decomposition of the ground state of the magneto-harmonic rotator.